\documentclass[aps,nofootinbib,manuscript]{revtex4}
\usepackage{bm}
\begin{document}

\title{The seed of magnetic monopoles in the early inflationary universe from a 5D vacuum state}
\author{ $^{1}$ Jes\'us Mart\'{\i}n Romero and $^{1,2}$ Mauricio
Bellini \footnote{E-mail address: mbellini@mdp.edu.ar}}

\address{$^{1}$ Departamento de F\'{\i}sica, Facultad de Ciencias Exactas y
Naturales, Universidad Nacional de Mar del Plata, Funes 3350,
(7600) Mar del Plata,
Argentina.\\
$^2$ Consejo Nacional de Investigaciones Cient\'{\i}ficas y
T\'ecnicas (CONICET). }

\begin{abstract}
Starting from a 5D Riemann flat metric, we have induced an
effective 4D Hermitian metric which has an antisymmetric part
which is purely imaginary. We have worked an example in which
both, non-metricity and cotorsion are zero. We obtained that the
production of monopoles should be insignificant at the end of
inflation  and the tensor metric should come asymptotically
diagonal and describing a nearly 4D de Sitter expansion.
\end{abstract}


\maketitle
\section{Introduction}

The possibility that our world may be embedded in a
$(4+d)$-dimensional universe with more than four large dimensions
has attracted the attention of a great number of researches. One
of these higher-dimensional theories, where the cylinder condition
of the Kaluza-Klein theory\cite{kk} is replaced by the conjecture
that the ordinary matter and fields are confined to a 4D subspace
usually referred to as a brane is the Randall and Sundrum
model\cite{rs}. The original version of the KK theory assures, as
a postulate, that the fifth dimension is compact. A few years ago,
a non-compactified approach to KK gravity, known as
Space-Time-Matter (STM) theory was proposed by Wesson and
collaborators\cite{im}. In this theory all classical physical
quantities, such as matter density and pressure, are susceptible
of a geometrical interpretation. Wesson's proposal also assumes
that the fundamental 5D space in which our usual spacetime is
embedded, should be a solution of the classical 5D vacuum Einstein
equations: $R_{AB}=0$\footnote{{In our conventions capital Latin
indices run from $0$ to $4$, greek indices run from $0$ to $3$ and
latin indices run from $1$ to $3$.}}. The mathematical basis of it
is the Campbell's theorem\cite{campbell}, which ensures an
embedding of 4D general relativity with sources in a 5D theory
whose field equations are apparently empty. That is, the Einstein
equations $G_{\alpha\beta}=-8\pi G\,T_{\alpha\beta}$ (we use
$c=\hbar=1$ units), are embedded perfectly in the Ricci-flat
equations $R_{AB}=0$.

The theory of magnetic monopoles was formulated many decades ago
by Dirac in \cite{Dir}. In his classical works, Dirac showed that
the existence of a magnetic monopole would explain the electric
charge quantization. Also, a lagrangian formulation describing the
electromagnetic interaction mediated by topologically massive
vector bosons-between charged, spin-1/2 fermions with an abelian
magnetic monopoles in a curved spacetime with nonminimally
coupling and torsion potential, was presented in\cite{prep}. In
the framework of inflation, magnetic monopole solutions of the
Einstein-Yang-Mill-Higgs equations with a positive cosmological
constant approach asymptotically the de Sitter spacetime
background and exist only for a nonzero Higgs
potential\cite{nucl}. More recently, Maxwell equations with
massive photons and magnetic monopoles were formulated using
spacetime algebra\cite{cafaro}. From the quantum-theoretical
standpoint monopoles and massive photons were discussed widely
during the last decades\cite{citt}. To discuss massive photons one
has to consider the Proca equations rather than the Maxwell ones.
Further, if a magnetic charge (monopole) really exists, the
Maxwell electrodynamics must be replaced by a generalized theory,
with the dual field tensor having a non vanishing divergence. Some
years ago the Weyl - Dirac formalism was generalized in order to
obtain a geometrically based general relativistic theory,
possessing electric and magnetic currents and admitting massive
photons\cite{isr}. In this letter we are interested to study the
evolution of primordial magnetic monopoles in the inflationary
epoch by extending Gravitoelectromagnetic Inflation and using some
ideas of Dirac\cite{dirac} and Einstein\cite{einstein} in the
framework of the Induced Matter theory, where the extra dimension
is space-like and noncompact. To make it we shall start from a 5D
vacuum state, on which we define null all the external sources and
magnetic monopoles [the absence of magnetic monopoles is a
characteristic of any ($N>4$) theory]. It means that the universe
in higher dimensions will be considered as empty. If we consider
the extra dimension as a complex function of other coordinates, it
is possible to obtain an effective 4D Hermitian tensor metric
which could be relevant to describe extended gravitation and
electrodynamics, where 4D magnetic monopoles are taken into
account. In this context, we shall study the production of
magnetic monopoles in the early inflationary universe.

\section{Formalism in a 5D vacuum}

We consider the 5D Riemann flat metric\cite{LB}
\begin{equation}\label{a1}
dS^2=\psi^2 dN^2-\psi^2\,e^{2N}\left(dr^2+r^2
d\Omega^2\right)-d\psi^2,
\end{equation}
with the action
\begin{equation}\label{a1'}
I=\int{d^4x
d\psi{\sqrt{\left|{{^{(5)}g}\over{^{(5)}g_0}}\right|}}\left[{{^{(5)}R}\over{16\pi
G}}+^{(5)}\mathcal{L}(A_B,A_{C;B})\right]},
\end{equation}
where ${^{(5)}g}=\psi^8 e^{6N} \,r^4\,{\rm sin}^2(\theta)$ is the
determinant of the covariant metric tensor $g_{AB}$, on a Riemman
spacetime. The metric (\ref{a1}) is Riemann-flat ($R_{ABCD}=0$)
and is a particular case of the so called canonical metric: $dS^2=
\psi^2 g_{\alpha\beta}(x^{\mu},\psi) dx^{\alpha}dx^{\beta} -
d\psi^2$, where ${\partial g_{\alpha\beta}\over\partial\psi}=0$,
so that $8\pi G \,G_{\alpha\beta}=-3g_{\alpha\beta}/\psi^2_0$,
$\Lambda = 3/(8\pi G \psi^2_0)$ being the cosmological constant.
Physically, this metric removes the potentials of electromagnetic
type and flattens the potential of scalar type, so that the fields
$A^B$ in the action (\ref{a1'}) must be considered as test ones.
Here,
${^{(5)}g_0}\equiv\left.{^{(5)}g}\right|_{[N=0,\psi=\psi_0,\theta=\pi/2]}$
is a constant. As in previous papers\cite{g1}, we shall consider a
Lagrangian density given by\footnote{We denote with comma ordinary
derivatives and with $(;)$ covariant derivatives.}
\begin{equation}\label{a1''}
^{(5)}\mathcal{L}(A_B,A_{C;B})=-{{1}\over{4}}{\cal Q}_{BC}{\cal
Q}^{BC},
\end{equation}
where ${\cal Q}^{AB}$ is an operator given by
\begin{equation}\label{a2}
{\cal Q}^{AB}= F^{AB}+\gamma\,\,g^{AB}\,A^{D}_{\,\,\,\,;D},
\end{equation}
such that $A_B=(A_\mu,\varphi)$, $A^B=(A^\mu,-\varphi)$, $\varphi$
being the inflaton field.

Since we are considering a 5D vacuum, one obtains absence of both,
5D gravitoelectric $J^B$ and gravitomagnetic $K^B$ currents
\begin{eqnarray}
&& \left({\cal Q}^{AB}\right)_{;A}=0, \\ \label{a3} && \left({\cal
Q}^{\dagger AB}\right)_{;A}=0 \label{a4},
\end{eqnarray}
where the dual tensor of ${\cal Q}^{AB}$ is
\begin{equation}
\left({\cal Q}^{\dagger}\right)^{
AB}={{1}\over{2}}{{\epsilon^{ABCD}}\over{\sqrt{\left|
^{(5)}g\right|}}}\,\,\, {\cal Q}_{CD}.
\end{equation}

\section{Induced gravitoelectromagnetic equations from a 5D vacuum}

Now we consider the 4D embedding $\psi\equiv \psi(r,N)$ on the 5D
metric (\ref{a1}), such that if we consider physical coordinates
$d\psi = {\partial\psi\over
\partial r} dr + {\partial\psi \over
\partial N} dN$. The induced operators ${\cal Q}^{\alpha\beta}$ will be
\begin{eqnarray}
&& {\cal Q}^{\alpha\beta}=
F^{\alpha\beta}+\gamma\,\,g^{\alpha\beta}\,\left(A^{\delta}\right)_{;\delta},
\label{a5} \\
&& \left({\cal Q}^{\alpha\beta}\right)_{;\alpha}=-4\pi\,\,J^\beta, \label{a3'} \\
&& \left({\cal Q}^{\dagger
\alpha\beta}\right)_{;\alpha}=-4\pi\,\,K^\beta, \label{a4'}
\end{eqnarray}
where $\left({\cal
Q}^{\dagger}\right)^{\alpha\beta}={{1}\over{2}}{{\epsilon^{\alpha\beta\mu\nu}}\over{\sqrt{\left|^{(4)}g\right|}}}\,\,\,{\cal
Q}_{\mu\nu}$. The induced gravitoelectric current is
\begin{equation}\label{a6}
J^\alpha=-{{1}\over{4\pi}}\left\{\left(F^{\alpha(\psi)}\right)_{;(\psi)}+\gamma\left[
g^{\alpha\beta}_{;\beta}A^{(\psi)}+
g^{\alpha(\psi)}_{;(\psi)}A^D_{\,\,;D}+g^{\alpha(\psi)}\left(A^D_{\,\,;D;(\psi)}+A^{(\psi)}_{\,\,;(\psi);\beta}\right)\right]\right\}.\end{equation}
The induced gravitomagnetic current is given by
\begin{equation}\label{a7}K^\beta=\left({\cal Q}^{\dagger
\alpha\beta}\right)_{;\alpha}=-{{1}\over{8\pi}}{{\epsilon^{\alpha\beta\mu\nu}}}
\left\{{{A_\sigma\left(\Gamma^{\sigma}_{\nu\mu}-\Gamma^{\sigma}_{\mu\nu}\right)+
\gamma
g_{\mu\nu}\Gamma^\eta_{\rho\eta}A^\rho}\over{\sqrt{\left|^{(4)}g\right|}}}\right\}_{;\alpha},
\end{equation}
where $\epsilon^{\alpha\beta\mu\nu}$ is the Ricci tensor density.
Notice that when torsion
$T^{\sigma}_{\,\,\,\nu\mu}={\Gamma^{\sigma}_{\,\,\,\mu\nu}-\Gamma^{\sigma}_{\,\,\,\nu\mu}\over
2}$ is zero and the metric tensor is diagonal, hence
gravitomagnetic current vanishes.

\section{Induced gravitoelectromagnetic equations on an Hermitic
metric}

In order to work an example with nonzero gravitomagnetic current
we shall study an example where the fifth coordinate $\psi \equiv
\psi(r,N)$ is a complex function of $N$ and $r$.

\subsection{An example with null cotorsion and non-metricity on an Hermitian metric}

We are interesting to study the case where the resultant 4D
effective metric is Hermitian
\begin{equation}\label{b1}
^{(4)}dS^2=\left[\left|\psi\right|^2+\left|{{\partial\psi}\over{\partial
N}}\right|^2\right] dN^2-\left[\left|\psi\right|^2  \,e^{2N}
+\left|{{\partial\psi}\over{\partial r}}\right|^2\right]
dr^2-{{\partial\psi}\over{\partial
N}}{{\partial\psi}\over{\partial r}}\left(dN dr- dr dN
\right)-\left|\psi\right|^2 e^{2N}r^2 d\Omega^2,
\end{equation}
such that $\left|\psi\right|^2 = \psi\,\psi^*$ and $U^{A} =
{dx^A\over dS}$ are the penta-velocities such that $g_{AB} U^A U^B
=1$. Furthermore, the conditions to obtain an Hermitian metric
tensor are:
\begin{equation}
\frac{\partial\psi}{\partial r} = \pm
\frac{\partial\psi^*}{\partial r}, \qquad
\frac{\partial\psi}{\partial N} = \mp
\frac{\partial\psi^*}{\partial N}.
\end{equation}
Hence, the metric (\ref{b1}) results to be represented by an
Hermitian tensor with $g_{\alpha\beta} = g^*_{\beta\alpha}$, where
the asterisk denotes the complex conjugate. The determinant
$g=det\left|g_{\alpha\beta}\right|$ is nonzero and real.
Furthermore, as in a real tensor metric, one has
$g_{\alpha\beta}\,g^{\gamma\beta}=\delta_{\beta}^{\,\,\,\gamma}$,
where $\delta_{\beta}^{\,\,\,\gamma}$ is the Kronecker tensor.
Here, the order of indices is important and, for example:
$g_{\alpha\beta}\,g^{\beta\gamma}\neq
\delta_{\beta}^{\,\,\,\gamma}$. Recently, Hermitian metrics has
been subject of interest to describe cosmology\cite{man}. The
effective 4D conections are given by
\begin{equation}\label{a12} \Gamma^{\lambda}_{\nu\mu}+g^{\lambda\sigma}\Gamma^{\kappa}_{\mu\sigma}g_{\left[\nu\kappa\right]}=
\left|^{\lambda}_{\mu\nu}\right|
+C^{\lambda}_{\mu\nu}+{{1}\over{2}}g^{\lambda\sigma}\left(Q_{\nu\mu\sigma}+Q_{\mu\nu\sigma}-Q_{\sigma\mu\nu}\right),\end{equation}
where the non-metricity terms are due to $Q_{\nu\mu\sigma}\equiv
-g_{\mu\sigma;\nu}$, torsion are related with cotorsion terms by
$2T^{\alpha}_{\mu\nu}=C^{\alpha}_{\mu\nu}-C^{\alpha}_{\nu\mu}$ and
\begin{equation}\label{chris2'}
\left|^{\lambda}_{\mu\nu}\right|={{1}\over{2}}\,\,g^{\lambda\sigma}{\left(g_{\mu\sigma,\nu}
+g_{\nu\sigma,\mu}-g_{\mu\nu,\sigma}\right)}.
\end{equation}
Finally, the non-zero gravitomagnetic currents components on the
effective 4D metric, are
\begin{eqnarray}
&& K^{N}=-\frac{\epsilon^{\alpha N
\mu\nu}}{4\pi}\left({{A_{\sigma}T^{\sigma}_{\mu\nu}}\over{\sqrt{\left|^{(4)}g\right|}}}\right)_{;\alpha},
\label{a8} \\
&& K^{r}=-\frac{\epsilon^{\alpha r
\mu\nu}}{4\pi}\left({{A_{\sigma}T^{\sigma}_{\mu\nu}}\over{\sqrt{\left|^{(4)}g\right|}}}\right)_{;\alpha},\label{a9}
\\
&& K^{\theta}={{1}\over{4\pi
\sqrt{\left|^{(4)}g\right|}}}\left\{2\left[\left(A_{\sigma}T^{\sigma}_{Nr}\right)_{;\varphi}-\left(A_{\sigma}T^{\sigma}_{N\varphi}\right)_{;r}-
\left(A_{\sigma}T^{\sigma}_{\varphi r}\right)_{;N}\right]
+\gamma\left(g_{\left[Nr\right]}\Gamma^{\eta}_{\rho\eta}A^{\rho}\right)_{;\varphi}\right\},\label{a10}
\\
&&
K^{\varphi}=-{{1}\over{4\pi\sqrt{\left|^{(4)}g\right|}}}\left\{2\left[\left(A_{\sigma}T^{\sigma}_{Nr}\right)_{;\theta}-
\left(A_{\sigma}T^{\sigma}_{N\theta}\right)_{;r}-\left(A_{\sigma}T^{\sigma}_{\theta
r}\right)_{;N}\right] +\gamma\left(g_{\left[Nr\right]}
\Gamma^{\eta}_{\rho\eta}A^{\rho}\right)_{;\theta}\right\}.\label{a11}
\end{eqnarray}

Now we consider the equation (\ref{a12}) with both, zero cotorsion
an non-metricity. In this case the last two terms in the right
side are zero, and we obtain
\begin{equation}\label{a12'}
\Gamma^{\lambda}_{\nu\mu}+g^{\lambda\sigma}\Gamma^{\kappa}_{\mu\sigma}g_{\left[\nu\kappa\right]}=
\left|^{\lambda}_{\mu\nu}\right|,
\end{equation}
and the gravitomagnetic currents for this particular case results
to be
\begin{eqnarray}
&& K^N=K^r=0,\label{ej1} \\
&& K^{\theta}=-{{\gamma
g_{[rN]}}\over{4\pi\sqrt{\left|^{(4)}g\right|}}}\left[\left({{g_{[Nr],\rho}}\over{g_{[Nr]}}}+\left\{{^{\theta}_{\rho\theta}}\right\}
+\left\{{^{\varphi}_{\rho\varphi}}\right\}\right)A^{\rho}\right]_{;\varphi},
\label{ej2} \\
&& K^{\varphi}=-{{\gamma
g_{[Nr]}}\over{4\pi\sqrt{\left|^{(4)}g\right|}}}\left[\left({{g_{[Nr],\rho}}
\over{g_{[Nr]}}}+\left\{{^{\theta}_{\rho\theta}}\right\}+\left\{{^{\varphi}_{\rho\varphi}}\right\}\right)A^{\rho}\right]_{;\theta},
\label{ej3}
\end{eqnarray}
where $g_{[\alpha\beta]}$ denotes the antisymmetric components of
the tensor metric and $\left\{{^{\alpha}_{\beta\gamma}}\right\}$
are the second kind Christoffel symbols. Notice that the existence
of non-zero gravitomagnetic current components depends of the
antisymmetry of the tensor metric. An interesting example, which
is relevant for cosmology is the case in which the complex
function $\psi(r,N)$ is given by
\begin{displaymath}
\psi(N,r)=\psi_0\,\left[e^{-r}+i\,e^{-N}\right],
\end{displaymath}
such that, at the end of inflation, the non-diagonal part of the
metric tensor becomes negligible with respect to the diagonal one.
If we take $\psi_0=1/H$ (when $H$ is the Hubble parameter, which
in this case is a constant) and $N=H t$, at the end of inflation
the asymptotic tensor metric is diagonal
\\\\
\begin{equation}\left.g_{\mu\nu}\right|_{t\gg 1/H}\simeq \left(%
\begin{array}{cccc}
  1 & 0 & 0 & 0 \\
  0 & -H^{-2} e^{2Ht}&  0 & 0 \\
  0 & 0 & -H^{-2} e^{2Ht} r^2 & 0\\
  0 & 0 & 0 & -H^{-2} e^{2Ht} r^2  sin^2(\theta)
\end{array}\right),%
\end{equation}
which describes a de Sitter expansion\cite{ultimo}. Notice that,
at the end of inflation $\left.U^{\psi}\right|_{t \gg 1/H}
\rightarrow 0$, and $\left.U^{t}\right|_{t \gg 1/H} \rightarrow 1$
for $U^{\theta}=U^{\phi}=0$, so that observers are at this time in
a nearly comoving frame: $\left.U^{r}\right|_{t\gg 1/H}
\rightarrow 0$.

\section{Final Remmarks}

In this letter we have extended Gravitoelectromagnetic Inflation
using some ideas of Dirac\cite{dirac} and Einstein\cite{einstein},
in the framework of the Induced Matter theory, where the extra
dimension is space-like, noncompact and complex. Starting from a
5D Riemann flat metric (\ref{a1}), we have induced an effective 4D
Hermitian metric which has an antisymmetric part $g_{[\mu\nu]}$
which is purely imaginary. We have worked an example where the
only non-diagonal components of the tensor metric are
$g_{Nr}=i\,{\partial\psi\over\partial N}
{\partial\psi\over\partial r}$ and
$g_{rN}=-i\,{\partial\psi\over\partial r}
{\partial\psi\over\partial N}$ in which both, non-metricity and
cotorsion are zero. In this case the non-zero components of the
gravitomagnetic current during inflation are $K^{\theta}$ and
$K^{\varphi}$. However, at the end of inflation the production of
monopoles should be insignificant and the tensor metric should
describe a nearly 4D de Sitter expansion. The evolution of the
universe from a de Sitter expansion to other
Friedmann-Robertson-Walker cosmology was studied using different
approaches\cite{seamb}.

\centerline{\bf{Acknowledgements}} \vskip .2cm The authors
acknowledge CONICET and UNMdP (Argentina) for financial support. \\

\end{document}